\documentstyle[12pt]{article}

\newcommand{\resection}[1]{\setcounter{equation}{0}\section{#1}}

\newcommand{\EQ}{\begin{equation}}
\newcommand{\EN}{\end{equation}}
\newcommand{\bea}{\begin{eqnarray}}
\newcommand{\eea}{\end{eqnarray}}
\newcommand{\hs}{\hspace{0.1cm}}

\newcommand{\be}{\beta}

\begin{document}
\setcounter{page}{0}
\topmargin 0pt
\oddsidemargin 5mm
\renewcommand{\thefootnote}{\fnsymbol{footnote}}
\newpage
\setcounter{page}{0}
\begin{titlepage}
\begin{flushright}
ISAS/EP/92/236\\
\end{flushright}
\vspace{0.5cm}
\begin{center}
{\large {\bf Correlation Functions in 2-Dimensional
Integrable Quantum Field Theories}} \\
\vspace{1.8cm}
{\large G. Mussardo}\\
\vspace{0.5cm}
{\em International School for Advanced Studies,\\
and \\
Istituto Nazionale di Fisica Nucleare\\
34014 Trieste, Italy}\\

\end{center}
\vspace{1.2cm}
\renewcommand{\thefootnote}{\arabic{footnote}}
\setcounter{footnote}{0}

\begin{abstract}
In this talk I discuss the form factor approach used to compute
correlation functions of integrable models in two dimensions. The
Sinh-Gordon model is our basic example. Using Watson's and the recursive
equations satisfied by matrix elements of local operators, I present
the computation of the form factors of the elementary field $\phi(x)$
and the stress-energy tensor $T_{\mu\nu}(x)$ of the theory.
\end{abstract}

\vspace{1.2cm}
\noindent

\end{titlepage}

\newpage

\resection{Introduction}

Important progress in Quantum Field Theories has been made in recent years
by studying two dimensional relativistic models. The reason essentially
lays in the successful application of several non-perturbative approaches
which lead to exact solutions of the quantum field dynamics. In this
talk I am concerned with the determination of multi-point correlation
functions of local operators for a massive field theory
\EQ
G_n(r_1,r_2,\ldots,r_n)\,=\,
<{\cal O}_{i_1}(r_1)\,{\cal O}_{i_2}(r_2)\ldots {\cal O}_{i_n}(r_n)>
\,\,\, .
\EN
My basic example will the the Sinh-Gordon model which is discussed
in a work made in collaboration with A. Fring and P. Simonetti \cite{FMS}.

For two-dimensional systems, there is usually a special limit
where the computation of the $G_n$'s is greatly simplified. This is the
ultraviolet asymptotic regime of the correlation functions, defined by
\[
|r_i-r_j|\ll \xi \,\,\, .
\]
where the correlation length is given by $\xi=m^{-1}$. This regime can be
equivalently reached by keeping finite values of the distances but taking
the limit $\xi\rightarrow\infty$. Either ways, a scale invariant behaviour
occurs in the model and the resulting 2-D infinite-dimensional conformal
symmetry induces a system of linear differential equations satisfied by the
correlators $G_n$ \cite{BPZ}. Their solution provides an explicit expression
for the correlation functions of the theory in the ultraviolet regime
\cite{BPZ,DF,ISZ}. However, the behaviour of correlation functions in the
cross-over region
\[
|r_i-r_j| \simeq \xi \,\, ,
\]
and in the infrared limit
\[
|r_i-r_j| \gg \xi \,\, ,
\]
is more complicated. Consider, for instance, the case of two-point
correlation functions
\EQ
<{\cal O}_i(r)\,{\cal O}_i(0)>\,=\,\frac{1}{\,\,r^{\eta_i}}\,
\Phi_i(mr) \,\,\, .
\EN
With this parametrization, the exponent $\eta_i$ is usually identified with
twice the anomalous dimension of the field ${\cal O}_i$ whereas $\Phi_i(x)$
is a scaling function\footnote{Scaling functions can be introduced as well for
the $n$-point correlators. We concentrate our discussion on the two-point
functions for clarity and simplicity.} of the variable $x=mr$. Although
conformal invariance severely restricts the possible values of $\eta_i$ and
gives quite powerful classification of the ultraviolet behaviours of
two-dimensional models, the computation of the scaling functions $\Phi_i(r)$
is, on the contrary, quite difficult.

There exists, however, a large class of relativistic models where the
determination of the scaling functions can be explicitly worked out. This is
the case of the massive integrable systems whose dynamics is strongly
constrained by an infinite number of integrals of motion. In particular,
the $S$-matrix of these models presents factorization and elasticity
properties and can be explicitly constructed [5-11]. In order to compute
multi-point correlators for massive integrable models, we may exploit their
spectral representation, i.e. their decomposition into an infinite sum over
intermediate multi-particle contributions (fig.\,1). For instance, the
two-point function of a hermitian scalar operator ${\cal O}_i(x)$ in real
Euclidean space can be written as\footnote{$\beta$ is the
rapidity variable, related to the two-dimensional momentum by
\EQ
p^0\,=\,m\cosh\beta \,\,\, ,\,\,\, p^1\,=\,m\sinh\beta\,\,\, .
\EN}
\begin{eqnarray}
& &\langle{\cal O}_i(x)\,{\cal O}_i(0)\rangle\,=
\sum_{n=0}^{\infty}
\int \frac{d\beta_1\ldots d\beta_n}{n! (2\pi)^n}
<0|{\cal O}_i(x)|\beta_1,\ldots,\beta_n>_{\rm in}{}_{\rm in}
<\beta_1,\ldots,\beta_n|{\cal O}_i(0)|0>
\label{correlation} \nonumber \\
& &\hspace{5mm} =\,\sum_{n=0}^{\infty}
\int \frac{d\beta_1\ldots d\beta_n}{n! (2\pi)^n}
\mid F_n^{{\cal O}_i}(\beta_1\ldots \beta_n)\mid^2
\exp \left(-mr\sum_{i=1}^n\cosh\beta_i
\right)\,\, ,
\label{infsum}
\end{eqnarray}
where $r$ denotes the radial distance $r=\sqrt{x_0^2 + x_1^2}$.
\begin{center}
\begin{picture}(400,150)
\thicklines
\put(150,70){\circle{40}}
\put(250,70){\circle{40}}
\put(150,70){\makebox(0,0){${\cal O}_i(x)$}}
\put(250,70){\makebox(0,0){${\cal O}_i(0)$}}
\put(170,70){\line(1,0){60}}
\put(200,80){\makebox(0,0){.}}
\put(200,75){\makebox(0,0){.}}
\put(200,60){\makebox(0,0){.}}
\put(200,55){\makebox(0,0){.}}
\put(150,90){\line(1,0){100}}
\put(150,50){\line(1,0){100}}
\end{picture}
\end{center}
\vspace{1mm}
\begin{center}
{\bf Figure 1}
\end{center}
The functions
\EQ
F_n^{{\cal O}_i}(\beta_1,\beta_2,\ldots,\beta_n)\,=\,
<0|{\cal O}_i(0)|\beta_1,\beta_2,\ldots,\beta_n>_{\rm in}
\EN
are the {\em form factors} of the operator ${\cal O}_i$ (fig.\,2).
\begin{center}
\begin{picture}(280,160)
\thicklines
\put(150,80){\circle{40}}
\put(130,80){\line(-1,0){30}}
\put(150,100){\line(0,1){30}}
\put(150,60){\line(0,-1){30}}
\put(150,80){\makebox(0,0){$\varphi_i(0)$}}
\put(140,97.33){\line(-1,1){27}}
\put(189,118){\makebox(0,0){$\beta_1$}}
\put(189,58){\makebox(0,0){$\beta_3$}}
\put(140,62.67){\line(-1,-1){28}}
\put(140,125){\makebox(0,0){$\beta_n$}}
\put(115,110){\makebox(0,0){$\beta_{n-1}$}}
\put(170,80){\line(1,0){30}}
\put(115,62){\makebox(0,0){.}}
\put(117,58.55){\makebox(0,0){.}}
\put(119,55.75){\makebox(0,0){.}}
\put(197,73){\makebox(0,0){$\beta_2$}}
\put(160,97.33){\line(1,1){25}}
\put(160,62.67){\line(1,-1){25}}
\end{picture}
\end{center}
\vspace{3mm}
\begin{center}
{\bf Figure 2}
\end{center}

\vspace{3mm}
\noindent
Analytic properties of the form factors have been investigated by several
authors and particularly important contributions can be found in [13-19]. As I
will show in the following, the form factor approach is a successful method in
order to compute correlation functions. The expansion of the correlators
in terms of the number of intermediate particles rather than in
a perturbative series of the coupling constant presents, in fact, several
advantages. First of all, the form factors take into account coupling constant
dependence of correlation functions to all order in perturbation theory.
Secondly, expressions like that one in eq.\,(\ref{infsum}) are fast convergent
series in the number $N$ of intermediate particles even for moderate value of
the distance and therefore, for practical applications, it is often
sufficient to compute the form factors of the operator ${\cal O}_i$ with
few external particles. Moreover, this spectral representation of
two-point functions formally coincides with the grand partition function
of a one-dimensional gas system, as noticed in \cite{Yurov,CMform}.
Hence, one can borrow successful techniques developped for one-dimensional
system in order to extract non-perturbative parameters of the theory. For
instance, as shown in \cite{Yurov,CMform}, the anomalous dimension of the
operator ${\cal O}$ in the ultraviolet limit is equal to the pressure of
the corresponding one-dimensional gas at one specific value of the fugacity.

I will focalize on the form factor approach for the
Sinh-Gordon model, described by the
Lagrangian
\EQ
{\cal L}\,=\,\frac{1}{2}(\partial_{\mu}\phi)^2\,-\,
\frac{m_0^2}{g^2} \cosh g\,\phi(x)\,\, .
\label{Sinh-Gordon}
\EN
The SG model belongs to the class of Toda Field Theory constructed on the
simply-laced root systems, in this case $A_1$. The reason why I
will consider this model is its simplicity which allows us to understand
the basic principles of the computation without being masked by algebraic
complicancies. Let us discuss initially the basic features of this theory.

\resection{The Sinh-Gordon Theory}

The SG model is the simplest example of an affine Toda Field Theories
\cite{Toda}, possessing a $Z_2$ symmetry $\phi\rightarrow -\phi$. By an
analytic continuation in $g$, i.e $g\rightarrow i g$, it can formally be
mapped to the Sine-Gordon model.

There are numerous alternative viewpoints for the Sinh-Gordon model.
First, it can be regarded either as a perturbation of the free massless
conformal action by means of the relevant operator\footnote{Although
the anomalous dimension of this operator (computed with
respect to the free conformal point), is negative, $\Delta=-g^2/8\pi$,
the resulting theory is unitary. This is due to the existence of non a
nonzero vacuum expectation values of some of the fields ${\cal O}_i$ in
the theory. A detailed discussion of this point can be found in \cite{YLZam}.}
$\cosh\,g\phi(x)$. Alternatively, it can be considered as a perturbation of
the conformal Liouville action
\EQ
{\cal S}\,=\,\int d^2x \left[
\frac{1}{2}(\partial_{\mu}\phi)^2-\lambda e^{g\phi} \right]
\label{Liouville}
\EN
by means of the relevant operator $e^{-g\phi}$ or as a conformal affine
$A_1$-Toda Theory \cite{Babon} in which the conformal symmetry is broken by
setting the free field to zero.

In a perturbative approach to the quantum field theory defined by the
Lagrangian (\ref{Sinh-Gordon}), the only ultraviolet divergences which
occur in any order in $g$ come from tadpole graphs and can be removed
by a normal ordering prescription with respect to an arbitrary mass scale
$M$. All other Feynman graphs are convergent and give rise to
finite wave function and mass renormalisation. The coupling constant $g$
does not renormalise.

An essential feature of the Sinh-Gordon theory is its integrability, which
in the classical case can be established by means of the inverse scattering
method \cite{Faddev}. In order to obtain the expressions of the (classical)
conserved currents one can use B\"{a}cklund transformation, as shown in
\cite{FMS}, and obtains an infinite set of conservation laws
\EQ
\partial_{z}\,T_{s+1}\,=\,
\partial_{\overline z}\,\Theta_{s-1}\,\, .
\EN
The corresponding charges ${\cal Q}_s$ are given by
\EQ
{\cal Q}_s\,=\,\oint \left[ T_{s+1}\,dz+\Theta_{s-1}\,d\overline z\right]
\,\,\, .
\EN
The integer-valued index $s$ which labels the integrals of motion is the spin
of the operators. Non-trivial conservation laws are obtained for odd values of
$s$
\EQ
s=1,3,5,7,\ldots
\label{conservedspin}
\EN
In analogy to the Sine-Gordon theory \cite{Sazaki}, an infinite set of
conserved charges ${\cal Q}_s$ with spin $s$ given in (\ref{conservedspin})
also exists for the quantized version of the Sinh-Gordon theory. They are
diagonalised by the asymptotic states with eigenvalues given by
\EQ
{\cal Q}_s \,|\beta_1,\ldots,\beta_n>\,=\,
\chi_s\,\sum_{i=1}^n e^{s\beta_i}\, |\beta_1,\ldots,\beta_n>\,\,\, ,
\label{charges}
\EN
where $\chi_s$ is the normalization constant of the charge ${\cal Q}_s$. The
existence of these higher integrals of motion precludes the possibility of
production processes and hence guarantees that the $n$-particle scattering
amplitudes are purely elastic and factorized into $n(n-1)/2$ two-particle
$S$-matrices. The exact expression for the Sinh-Gordon theory is given by
\cite{AFZ}
\EQ
S(\beta,B)\,=\,
\frac{\tanh\frac{1}{2}(\beta-i\frac{\pi B}{2})}
{\tanh\frac{1}{2}(\beta+i\frac{\pi B}{2})} \label{smatrix}\,\, ,
\EN
where $B$ is the following function of the coupling constant $g$
\EQ
B(g)\,=\,\frac{2g^2}{8\pi+g^2} \hs\hs\hs.
\EN
This formula has been checked against perturbation theory in ref.\,\cite{AFZ}
and can also be obtained by analytic continuation of the $S$-matrix of the
first breather of the Sine-Gordon theory \cite{ZZ}. For real values of $g$ the
$S$-matrix has no poles in the physical sheet and hence there are no bound
states, whereas two zeros are present at the crossing symmetric positions
\EQ
\beta\,=\,
\left\{
\begin{array}{l}
i\frac{\pi B}{2} \\
i\frac{\pi (2-B)}{2}
\end{array}
\right.
\EN
An interesting feature of the $S$-matrix is its invariance under the map
\cite{MC}
\EQ
B\rightarrow 2-B \label{mapd}
\EN
{\em i.e.} under the {\em strong-weak} coupling constant duality
\EQ
g\rightarrow \frac{8\pi}{g} .\label{mapdu}
\EN
This duality is a property shared by the unperturbed conformal Liouville theory
(\ref{Liouville}) \cite{Mansfield} and it is quite remarkable that it survives
even when the conformal symmetry is broken.

\resection{Form Factors}

The form factors (FF) are matrix elements of local operators between the
vacuum and $n$-particle in-state
\EQ
F_n^{\cal O}(\beta_1,\beta_2,\ldots,\beta_n)\,=\,
<0|{\cal O}(0)|\beta_1,\beta_2,\ldots,\beta_n>_{\rm in}\,\, .
\EN
For local scalar operators ${\cal O}(x)$, relativistic invariance implies
that $F_n$ are functions of the difference of the rapidities. Except for the
poles corresponding to the one-particle bound states in all sub-channels, we
expect the form factors $F_n$ to be analytic inside the strip
$0 < {\rm Im } \be_{ij} < 2\pi$.

The form factors of a hermitian local scalar operator ${\cal O}(x)$ satisfy
a set of functional equations, known as Watson's equations \cite{Watson},
which for integrable systems assume a particularly simple form
\bea
F_n(\be_1, \dots ,\be_i, \be_{i+1}, \dots, \be_n) &=& F_n(\be_1,\dots,\be_{i+1}
,\be_i ,\dots, \be_n) S(\beta_i-\beta_{i+1}) \,\, ,
\label{permu1}\\
F_n(\be_1+2 \pi i, \dots, \be_{n-1}, \be_n ) &=& F_n(\be_2 ,\ldots,\be_n,
\be_1) = \prod_{i=2}^{n} S(\beta_i-\beta_1) F_n(\be_1, \dots, \be_n)
\,\, .
\nonumber
\eea
The first equation states that as a result of the commutation of two
particles in the asymptotic state we get a scattering process whereas
the second equation fixes the discontinuity of the functions $F_n$ on the
cuts $\beta_{1i} = 2 \pi i$. In the case $n=2$, eqs.\,(\ref{permu1}) reduce to
\EQ
\begin{array}{ccl}
F_2(\beta)&=&F_2(-\beta)S_2(\beta) \,\, ,\\
F_2(i\pi-\beta)&=&F_2(i\pi+\beta) \,\,\, .
\end{array}
\label{F2}
\EN
The general solution of Watson's equations for diagonal $S$-matrix systems
can always be brought into the form
\cite{Karowski}
\EQ
F_n(\beta_1,\dots,\beta_n) =K_n(\beta_1,\dots,\beta_n) \prod_{i<j}F_{\rm min}
(\beta_{ij})  \,\, ,
\label{parametrization}
\EN
where $F_{\rm min}(\beta)$ has the properties that it satisfies (\ref{F2}), is
analytic in $0\leq$ Im $\beta\leq \pi$, has no zeros in $0<$ Im $\beta<\pi$,
and converges to a constant value for large values of $\beta$. These
requirements uniquely determine this function, up to a normalization.
In the case of the SG model, $F_{\rm min}(\beta)$ is given by
\EQ
F_{\rm min}(\beta,B)\,=\,{\cal N}\,\exp\,\left[
8\int_0^{\infty} \frac{dx}{x} \frac{\sinh\left(\frac{x B}{4}\right)
\sinh\left(\frac{x}{2}(1-\frac{B}{2})\right) \,\sinh\frac{x}{2}}{\sinh^2 x}
\sin^2\left(\frac{x\hat\beta}{2\pi}\right)\right] \,\,\,.
\label{integral}
\EN
We choose our normalization to be
\EQ
{\cal N} \,=\,\exp\left[-4\int_0^{\infty}
\frac{dx}{x} \frac{\sinh\left(\frac{x B}{4}\right)
\sinh\left(\frac{x}{2}(1-\frac{B}{2})\right) \,\sinh\frac{x}{2}}{\sinh^2 x}
\right] \,\,\,.
\EN
The analytic structure of $F_{\rm min}(\beta,B)$ can be easily read from
its infinite product representation in terms of $\Gamma$ functions
\EQ
F_{\rm min}(\beta,B)\,=\,
\prod_{k=0}^{\infty}
\left|
\frac{\Gamma\left(k+\frac{3}{2}+\frac{i\hat\beta}{2\pi}\right)
\Gamma\left(k+\frac{1}{2}+\frac{B}{4}+\frac{i\hat\beta}{2\pi}\right)
\Gamma\left(k+1-\frac{B}{4}+\frac{i\hat\beta}{2\pi}\right)}
{\Gamma\left(k+\frac{1}{2}+\frac{i\hat\beta}{2\pi}\right)
\Gamma\left(k+\frac{3}{2}-\frac{B}{4}+\frac{i\hat\beta}{2\pi}\right)
\Gamma\left(k+1+\frac{B}{4}+\frac{i\hat\beta}{2\pi}\right)}
\right|^2
\EN
$F_{\rm min}(\beta,B)$ has a simple zero at the threshold $\beta=0$ since
$S(0)=-1$ and its asymptotic behaviour is given by
\EQ
\lim_{\beta \rightarrow \infty} F_{\rm min}(\beta,B) = 1.
\EN
It satisfies the functional equation
\EQ
F_{\rm min}(i\pi+\beta,B) F_{\rm min}(\beta,B)\,=\,
\frac{\sinh\beta}{\sinh\beta+\sinh\frac{i\pi B}{2}}
\label{shift}
\EN
which is useful in order to find a convenient form for the recursive equations
of the form factors.

A useful expression for the numerical evaluation of $F_{\rm min}(\beta,B)$
is given by
\begin{eqnarray}
& &F_{\rm min}(\beta,B) \,=\, {\cal N}
\prod_{k=0}^{N-1}
\left[\frac{\left(1+\left(\frac{\hat\beta/2 \pi}{k+\frac{1}{2}}\right)^2\right)
\left(1+\left(\frac{\hat\beta/2 \pi}{k+\frac{3}{2}-\frac{B}{4}}\right)^2\right)
\left(1+\left(\frac{\hat\beta/2 \pi}{k+1+\frac{B}{4}}\right)^2\right)}
{\left(1+\left(\frac{\hat\beta/2 \pi}{k+\frac{3}{2}}\right)^2\right)
\left(1+\left(\frac{\hat\beta/2 \pi}{k+\frac{1}{2}+\frac{B}{4}}\right)^2\right)
\left(1+\left(\frac{\hat\beta/2 \pi}{k+1-\frac{B}{4}}\right)^2\right)}\right]^
{k+1} \\
& &\times \,
\exp\,\left[
8\int_0^{\infty} \frac{dx}{x} \frac{\sinh\left(\frac{x B}{4}\right)
\sinh\left(\frac{x}{2}(1-\frac{B}{2})\right) \,\sinh\frac{x}{2}}{\sinh^2 x}
(N+1-N\,e^{-2x})\,e^{-2Nx}\,
\sin^2\left(\frac{x\hat\beta}{2\pi}\right)\right] \,\,\,.\nonumber
\end{eqnarray}
The rate of convergence of the integral may be improved substantially
by increasing the value of $N$.

The remaining factors $K_n$ in (\ref{parametrization}) then satisfy Watson's
equations with $S_2=1$, which implies that they are completely symmetric,
$2 \pi i$-periodic functions of the $\beta_{i}$. They must contain all the
physical poles expected in the form factor under consideration and must
satisfy a correct asymptotic behaviour for large value of $\beta_i$. Both
requirements depend on the nature of the theory and on the operator $\cal O$.

Taking into account the one-particle pole in the three-particle channel
at $\beta_{ij}=i\pi$, the general form factors of a scalar hermitian
operator in the Sinh-Gordon model can be parameterized as
\EQ
F_n(\beta_1,\ldots,\beta_n)\,=
H_n
\, Q_n(x_1,\ldots,x_n)\,
\prod_{i<j} \frac{F_{\rm min}(\beta_{ij})}{(x_i+x_j)}
\,\, , \label{para}
\EN
where we have introduced the variables
\EQ
x_i\,=\,e^{\beta_i} \,\, ,
\EN
and the normalization constant $H_n$. The functions $Q_n(x_1,\dots,x_n)$
are symmetric polynomials\footnote{The polynomial nature of the functions
$Q_n$ is dictated by the locality of the theory \cite{nankai}.} in the
variables $x_i$. They can be expressed in terms of {\em elementary symmetric
polynomial} $\sigma^{(n)}_k(x_1,\dots,x_n)$ which are generated by
\cite{Macdon}
\EQ
\prod_{i=1}^n(x+x_i)\,=\,
\sum_{k=0}^n x^{n-k} \,\sigma_k^{(n)}(x_1,x_2,\ldots,x_n).
\label{generating}
\EN

Relativistic invariance demands that the total degree of $Q_n$ entering a
form factor of spinless operators should be $n(n-1)/2$ (in order to match
the total degree of the denominator in (\ref{para})). The order of the degree
of $Q_n$ in each variable $x_i$ is fixed, on the other hand, by the nature and
by the asymptotic behaviour of the operator $\cal O$ which is considered.
We will come back to this point later, when we discuss specific operators.

\subsection{Pole Structure and Residue Equations for the Form Factors}

The pole structure of the form factors induces a set of recursive equations
for the $F_n$. For the Sinh-Gordon model there is only one kind of poles that
arises from kinematical poles located at $\beta_{ij}=i\pi$. They are related
to the one-particle pole in a subchannel of three-particle states which, in
turn, corresponds to a crossing process of the elastic $S$-matrix. The
corresponding residues are computed by the LSZ reduction
\cite{Smirnov1,Smirnov2} and give rise to a recursive equation
between the $n$-particle and the $(n+2)$-particle form factors (fig.\,3)
\EQ
-i\lim_{\tilde\beta \rightarrow \beta}
(\tilde\beta - \beta)
F_{n+2}(\tilde\beta+i\pi,\beta,\beta_1,\beta_2,\ldots,\beta_n)=
\left(1-\prod_{i=1}^n S(\beta-\beta_i)\right)\,
F_n(\beta_1,\ldots,\beta_n)  . \label{recursive}
\EN
\begin{center}
\begin{picture}(280,120)
\thicklines
\put(-30,0){\circle{40}}
\put(-50,0){\line(-1,0){30}}
\put(-30,20){\line(0,1){30}}
\put(-30,-20){\line(0,-1){30}}
\put(-30,0){\makebox(0,0){${\cal F}_n$}}
\put(-40,17.33){\line(-1,1){27}}
\put(-40,-17.33){\line(-1,-1){28}}
\put(-10,0){\line(1,0){30}}
\put(-65,-18){\makebox(0,0){.}}
\put(-63,-21.45){\makebox(0,0){.}}
\put(-61,-24.25){\makebox(0,0){.}}
\put(-20,17.33){\line(1,1){25}}
\put(-20,-17.33){\line(1,-1){25}}
\put(70,0){\vector(1,0){40}}
\put(200,0){\circle{40}}
\put(200,0){\makebox(0,0){${\cal F}_{n-2}$}}
\put(210,-17.33){\line(1,-1){25}}
\put(200,20){\line(0,1){30}}
\put(200,-20){\line(0,-1){30}}
\put(180,0){\line(-1,0){30}}
\put(165,-18){\makebox(0,0){.}}
\put(167,-21.45){\makebox(0,0){.}}
\put(169,-24.25){\makebox(0,0){.}}
\put(190,17.33){\line(-1,1){25}}
\put(190,-17.33){\line(-1,-1){25}}
\put(220,0){\line(1,0){45}}
\put(270,0){\circle{10}}
\put(275,0){\line(1,0){30}}
\put(267.5,4.331){\line(-1,1){30}}
\put(272.5,-4.331){\line(1,-1){30}}
\end{picture}
\end{center}
\vspace{5mm}
\begin{center}
{\bf Figure 3}
\end{center}
This equation establishes a recursive structure between the $(n+2)$- and
$n$-particle form factors.

\resection{Solution of the Recursive Equations}

According to the $Z_2$ symmetry of the SG model, we can label the operators
by their parity. For operators which are $Z_2$-odd the only possible non-zero
form factors are those involving an odd number of particles. For $Z_2$-even
operators the only possible non-zero form factors are those involving an even
number of particles. The vacuum expectation value of $Z_2$-even operators can
in principle be different from zero.

The simplest representative of the odd sector is given by the (renormalised)
field $\phi(x)$ itself. It creates a one-particle state from the vacuum.
Our normalization is fixed to be
\EQ
F_1^{\phi}(\beta)\,=\, <0\mid \phi(0) \mid \beta>_{\rm in}\,=\,
\frac{1}{\sqrt{2}} \,\, .
\label{normphi}
\EN
For the even sector, an important operator is given by the energy-momentum
tensor
\EQ
T_{\mu\nu}(x)\,=\,2\pi\,(:\partial_{\mu}\phi \partial_{\nu}\phi
-g_{\mu\nu} {\cal L}(x)\,:)
\EN
where $:\,:$ denotes the usual normal ordering prescription with respect
an arbitrary mass scale $M$. Its trace $T_{\mu}^{\,\mu}(x)=\Theta(x)$ is a
spinless operator whose normalization is fixed in terms of its two-particle
form factor
\EQ
F_2^{\Theta}(\beta_{12}=i\pi)\,=\, {}_{\rm out}<\beta_1\mid
\Theta(0) \mid \beta_2>_{\rm in}\,=\,2 \pi m^2
\,\, ,
\label{normtheta}
\EN
where $m$ is the physical mass.

In the following we shall compute the form factors of the operators $\phi(x)$
and $\Theta(x)$. Employing the parameterization (\ref{para}), the recursive
equations (\ref{recursive}) take on the form
\EQ
(-)^n\,Q_{n+2}(-x,x,x_1,\ldots,x_n)\, = \,x  D_n(x,x_1,x_2,\ldots ,x_n)
\,Q_n(x_1,x_2,\ldots,x_n)
\label{gleich}
\EN
where we have introduced the function
\EQ
D_n=\, {{-i} \over {4\sin(\pi B/2)}} \left(\prod_{i=1}^n
\left[(x+\omega x_i)(x-\omega^{-1}x_i)\right] - \prod_{i=1}^n \left[
(x-\omega x_i)(x+\omega^{-1}x_i)\right]\right)
\EN
with $\omega=\exp(i\pi B/2)$. The normalization constants for the
form factors of odd and even operators are conveniently chosen to be
\bea
H_{2n+1} &=& H_1 \left(\frac{4\sin(\pi B/2)}{ F_{\rm min}(i \pi,B)}
\right)^n \\
H_{2n} &=& H_2 \left(\frac{4\sin(\pi B/2)}{ F_{\rm min}(i \pi,B)}
\right)^{n-1}\nonumber
\eea
where $H_1$ and $H_2$ are the initial conditions, fixed by the nature
of the operator. Using the generating function (\ref{generating}) of the
symmetric polynomials, the function $D_n$ can be expressed as
\EQ
D_n={{1} \over {2 \sin(\pi B/2)}} \sum_{l,k=0}^n (-1)^l \sin\left(
(k-l) {{\pi B} \over{2}} \right) x^{2n -l-k} \sigma_l^{(n)} \sigma_k^{(n)}.
\label{sum}
\EN
As function of $B$, $D_n$ is invariant under $B\rightarrow -B$.

As shown in \cite{FMS}, the symmetric polynomials $Q_{2n+1}$
entering the form factors of the elementary field $\phi(x)$ can be factorized
as
\EQ
Q_{2n+1}(x_1,\ldots,x_{2n+1})\,=\,\sigma_{2n+1}^{(2n+1)} \,
P_{2n+1}(x_1,\ldots,x_{2n+1}) \hspace{5mm} n>0 \,\,\, ,
\label{field}
\EN
whereas the analogous polynomials entering the form factors of the trace of the
stress-energy tensor can be written as
\EQ
Q_{2n}(x_1,\ldots,x_{2n})\,=\,\sigma_1^{(2n)} \sigma_{2n-1}^{(2n)} \,
P_{2n}(x_1,\ldots,x_{2n}) \hspace{5mm} n>1 \,\,\, .
\label{trace}
\EN
$P_n(x_1,\ldots,x_n)$ is a symmetric polynomial of total degree $n(n-3)/2$
and of degree $n-3$ in each variable $x_i$. Using the following property of
the elementary symmetric polynomials
\EQ
\sigma_k^{(n+2)}(-x,x,x_1,\ldots,x_n)\,=\,
\sigma_k^{(n)}(x_1,x_2,\ldots,x_n)-x^2 \sigma_{k-2}^
{(n)}(x_1,x_2,\ldots,x_n)  \,\,\, ,
\label{kinshift}
\EN
the recursive equations (\ref{gleich}) can then be written in terms of the
$P_n$ as
\EQ
(-)^{n+1}\,P_{n+2}(-x,x,x_1,\ldots,x_n)\, = \,
\frac{1}{x} D_n(x,x_1,x_2,\ldots ,x_n)
\,P_n(x_1,x_2,\ldots,x_n) \,\, .
\label{gleich1}
\EN
Using the recursive equations (\ref{gleich1}) and the transformation
property of the elementary symmetric polynomials (\ref{kinshift}), the
explicit expressions of the first polynomials $P_n(x_1,\ldots,x_n)$ are
given by\footnote{The upper index of the elementary symmetric
polynomials entering $P_n$ is equal to $n$ and we suppress it, in order to
simplify the notation.}
\bea
P_3(x_1,\ldots,x_3)         &=& 1 \nonumber  \\
P_4(x_1,\ldots,x_4) &=& \sigma_2\nonumber \\
P_5(x_1,\ldots,x_5) &=&  \sigma_2 \sigma_3 -c_1^2
                              \sigma_5 \\
P_6(x_1,\ldots,x_6) &=& \sigma_3 (\sigma_2\sigma_4 -\sigma_6)  -c_1^2
(\sigma_4 \sigma_5 + \sigma_1 \sigma_2 \sigma_6 - \sigma_3 \sigma_6)
\nonumber \\
P_7(x_1,\ldots,x_7) &=& \sigma_2 \sigma_3 \sigma_4  \sigma_5  -c_1^2
(\sigma_4 \sigma_5^2 +\sigma_1 \sigma_2\sigma_5\sigma_6+
\sigma_2^2\sigma_3\sigma_7-c_1^2 \sigma_2\sigma_5\sigma_7)+ \nonumber\\
&& -c_2 (\sigma_1\sigma_2\sigma_4\sigma_7+
\sigma_3\sigma_5\sigma_6 -c_2 \sigma_1\sigma_6\sigma_7)+c_1^2 c_2^2
\sigma_7^2 \nonumber
\eea
where $c_1=2 \cos (\pi B/2)$ and  $c_2 = 1 - c_1^2$. Expression of the higher
$P_n$ are easily computed by an iterative use of eqs.\,(\ref{gleich}).
For practical application the first representatives of $P_n$ are sufficient to
compute with a high degree of accuracy the correlation functions of the fields.
In fact, the $n$-particle term appearing in the correlation function of the
fields (\ref{correlation}) behaves as $e^{-n(mr)}$ and for quite large
values of $mr$ the correlator is dominated by the lowest number of particle
terms. This conclusion is also confirmed by an application of the $c$-theorem
which I discuss at the end of the talk. It is interesting to notice that
closed expressions for $P_n$ can be found for particular values of the
coupling constant.

\subsubsection{The Self-Dual Point}

\,

The self-dual point in the coupling constant manifold has the special value
\EQ
B\left( \sqrt{8 \pi}\right) \, =\, 1 \,\, .
\EN
The two zeros of the $S$-matrix merge together and the function
$D_n(x,x_1,x_2,\ldots,x_n)$ acquires the particularly simple form
\EQ
D_n(x|x_1,x_2\ldots,x_n)\,=\,
\left(\sum_{k=0}^n (-1)^{k+1} \sin\frac{k\pi}{2} x^{n-k} \sigma_k^{(n)}
\right)\,
\left(\sum_{l=0}^n (-1)^{l} \cos\frac{l\pi}{2} x^{n-l} \sigma_l^{(n)}
\right)\,\hs\hs\hs. \label{recfun}
\EN
In this case the general solution of the recursive equations
(\ref{gleich1}) is given by \cite{FMS}
\EQ
P_n(x_1,x_2,\dots,x_n) \,=\,
{\rm det}\, {\cal A}(x_1,x_2,\ldots,x_n) \label{loesung}
\EN
where ${\cal A}$ is an $(n-3)\times (n-3)$ matrix whose entries are
\EQ
{\cal A}_{ij}(x_1,x_2,\ldots,x_n)\,=\,
\sigma^{(n)}_{2j-i+1}\,\cos^2\left[(i-j)\frac{\pi}{2}
\right] \,\,\, ,
\EN
i.e.
\EQ
{\cal A} = \left( \begin{array}{lllll}
\sigma_2 &    0     & \sigma_6 &   0      & \cdots \\
   0     & \sigma_3 &    0     & \sigma_7 & \cdots \\
   1     &    0     & \sigma_4 &   0      & \cdots \\
   0     & \sigma_1 &    0     & \sigma_5 & \cdots \\
 \vdots  &  \vdots  &  \vdots  &  \vdots  & \ddots \\
\end{array} \right)
\EN
This can be proved by exploiting the properties of determinants.

\subsubsection{The ``Inverse Yang-Lee" Point}

\,

A closed solution of the recursive equations (\ref{gleich1}) is also obtained
for
\EQ
B\left( 2 \sqrt{ \pi} \right) = { 2 \over 3} \,\,\, .
\EN
The reason is that, for this particular value of the coupling constant
the S-matrix of the Sinh-Gordon theory coincides with the inverse of the
$S$-matrix $S_{\rm YL}(\beta)$ of the Yang-Lee model \cite{Cardymus} or,
equivalently
\EQ
S(\beta,-\frac{2}{3})\,=\,S_{\rm YL}(\beta)\,\,\, .
\EN
Since the recursive equations (\ref{gleich1}) are invariant under
$B\rightarrow -B$ (see sect.\,4.2), a solution is provided by the same
combination of symmetric polynomials found for the Yang-Lee model
\cite{Smirnov2,YLZam}, i.e.
\EQ
P_n(x_1,x_2,\dots,x_n) \, = \,
{\rm det}\, {\cal B}(x_1,x_2,\ldots,x_n) \label{loesungyl}
\EN
with the following entries of the $(n-3) \times (n-3)$-matrix ${\cal B}$
\EQ
{\cal B}_{ij}\, = \, \sigma_{3j-2i+1} \,\,\, .
\EN

\resection{Form factors and $c$-theorem}

\,

The Sinh-Gordon model can be regarded as deformation of the free massless
theory with central charge $c=1$. This fixed point governs the ultraviolet
behaviour of the model whereas the infrared behaviour corresponds to a massive
field theory with central charge $c=0$. Going from the short- to
large-distances, the variation of the central charge is dictated by the
$c$-theorem of Zamolodchikov \cite{cth}. An integral version of this theorem
has been derived by Cardy \cite{Cardycth} and related to the spectral
representation of the two-point function of the trace of the stress-energy
tensor in \cite{Friedan,Freedman}, i.e.
\EQ
\Delta c \,=\, \int_0^{\infty} d\mu\, c_1(\mu)\,\, ,
\label{variation}
\EN
where $c_1(\mu)$ is given by
\begin{eqnarray}
& & c_1(\mu)\,=\,\frac{6}{\pi^2}\frac{1}{\mu^3} {\rm Im}\, G(p^2=-\mu^2)
\,\, ,
\label{spectral} \\
& & G(p^2) \,=\, \int d^2 x\, e^{-ip\dot x} \,
<0|\Theta(x)\Theta(0)|0>_{\rm conn}
\nonumber \,\,\, .
\end{eqnarray}
Inserting a complete set of in-state into (\ref{spectral}), we can express
the function $c_1(\mu)$ in terms of the form factors $F_{2n}^{\Theta}$
\begin{eqnarray}
c_1(\mu)& =&\frac{12}{\mu^3} \sum_{n=1}^{\infty} \frac{1}{(2n)!}
\int\frac{d\beta_1\ldots d\beta_{2n}}{(2\pi)^{2n}}\,
\mid F_{2n}^{\Theta}(\beta_1,\ldots, \beta_{2n})\mid^2 \\
& & \,\,\, \times \,
\delta(\sum_i m\sinh\beta_i)\,\delta(\sum_i m\cosh\beta_i-\mu)\,\,\, .
\nonumber
\end{eqnarray}
For the Sinh-Gordon theory $\Delta c=1$ and it is interesting to study
the convergence of this series increasing the number of intermediate
particles. For the two-particle contribution, we have the following
expression
\EQ
\Delta c^{(2)}\,=\,\frac{3}{2 F^2_{\rm min}(i\pi)}
\,\int_{0}^{\infty} \frac{d\beta}{\cosh^4\beta} \,|F_{\rm min}(2\beta)|^2
\,\,\, .
\EN
The numerical results for different values of the coupling
constant $g^2/4\pi$ are listed in the table below
\begin{table}

$$ \begin{array}{|c|c|c|}
 \hline
& &  \\
B
& \frac{g^2}{4\pi}
& \Delta\,c^{(2)}
\\
   &        &        \\
\hline
   &        &        \\
\frac{1}{500} & \frac{2}{999} & 0.9999995 \\
\frac{1}{100} & \frac{2}{199} & 0.9999878 \\
\frac{1}{10} & \frac{2}{19} &   0.9989538  \\
\frac{3}{10} & \frac{6}{17} &   0.9931954  \\
\frac{2}{5} & \frac{1}{2} &     0.9897087  \\
\frac{1}{2} & \frac{2}{3} &     0.9863354  \\
\frac{2}{3} & 1 &     0.9815944  \\
\frac{7}{10} & \frac{14}{13} &     0.9808312  \\
\frac{4}{5} & \frac{4}{3} &     0.9789824  \\
1 & 2 &     0.9774634  \\
& & \\ \hline
\end{array}
$$
\end{table}
It is evident that the sum rule is saturated by the two-particle form factor
also for large values of the coupling constant.  Hence, the expansion in the
number of intermediate particles results in a fast convergent series, as it is
confirmed by the computation of the next terms involving the form factor with
four and six particles.

\resection{Conclusions}

\,

The computation of the Green functions is a central problem in a Quantum Field
Theory. For integrable models, a promising approach to this question is
given by the bootstrap principle applied to the computation of the matrix
elements of local operators. It would be interesting to use this approach
in order to derive differential equations satisfied by the quantum correlators
and also to classify the operator content of a quantum integrable field theory.
\vspace{1cm}

\section*{Acknowledgments}

I would like to thank A. Fring and P. Simonetti for our collaboration on
this project and S. Elitzur and A. Schwimmer for useful discussions.

\end{document}